Lukas R.A. Wilde

# KI-Bilder und die Widerständigkeit der Medienkonvergenz: Von primärer zu sekundärer Intermedialität?

Zum Zeitpunkt des Verfassens dieses Textes – im Frühjahr 2024 – existieren allgemein zugängliche KI-Bildgeneratoren wie Dall·E, Midjourney oder Stable Diffusion seit etwa zwei Jahren. Neben diesen vermutlich bekanntesten Anbietern finden sich zahlreiche weitere Alterativen zur KI-Bilderzeugung wie etwa Ideogram, Leonardo.AI, Gemini, Blue Willow oder Firefly; beinahe im Wochentakt werden neue Anwendungen auf den Online-Markt gebracht oder angekündigt, zuletzt mit OpenAIs Sora sogar Bewegtbildgeneratoren. Ziemlich genau in der Halbzeit dieser bisherigen Phase, also im Frühjahr 2023, wurde mit der *IMAGE*-Sonderausgabe *Generative Imagery: Towards a ›New Paradigm‹ of Machine Learning-Based Image Production* (an der der Verfasser dieses Textes ebenfalls als Autor und Mitherausgeber beteiligt war) eine erste medienwissenschaftliche und bildtheoretische Bestandsaufnahme oder ›Vermessung‹ dieses neuen Paradigmas der Bild-Produktion, -distribution und -rezeption vorgelegt. Die meisten der damals von den beteiligten Autor*innen getroffenen Befunde waren ohne Frage nur wenige Monate später bereits technisch überholt, etwa Beschreibungen der damals typischen, technisch bedingten Grenzen bzw. ›Bugs und Glitches‹ der neuen Bildtechnik. Die häufig falsche Anzahl an menschlichen Fingern (vgl. WASIELEWSKI 2023) war in Midjourney V.5, veröffentlicht im März 2023, etwa weitgehend gelöst. Waren text-to-image-Generatoren und text-to-text-Generatoren (wie heute ChatGPT, Claude oder Gemini) im Frühjahr 2022 zudem noch technisch strikt geschieden, veränderte ChatGPT-3 die KI-Bildproduktion im darauffolgenden Oktober mit der Integration von Dall·E3 abermals grundlegend. Seither verfügt die iterative Bildgeneration auch über ein dialogisches Gedächtnis. Sicherlich wird auch der vorliegende Text zum Zeitpunkt seiner Publikation in vielfacher Hinsicht bereits hoffnungslos veraltet sein. Während er verfasst wurde, begannen etwa die ersten KI-Kurzvideos von OpenAIs Sora zu zirkulieren, die KI-Bilder in eine Zeitdimension hinein verlängern und den Phänomen- und Beschreibungsebenen wohl abermals eine neue Richtung vorgeben werden.

Für wissenschaftliche Publikationszyklen stellt dies ein gewaltiges Problem dar, möchte man einerseits nicht nur stets veraltete, zum Zeitpunkt ihres Erscheinens bereits historische Momentaufnahmen in die Welt entlassen, andererseits aber auch nicht über eine Zukunft spekulieren, die in KI-Fragen ganz besonders von Marketing-Utopien und imaginierten Techno-Katastrophen vernebelt wird. Die hier folgenden Reflexionen nehmen daher eine wesentlich größere Latenz zum Ausgangspunkt, die vermutlich auch in ein, zwei Jahren noch nicht gänzlich überwunden und entschieden sein wird. Gefragt wird nach der Integration bzw. Widerständigkeit von KI-Bildern innerhalb der weiteren Medienkonvergenz oder der Medienökologie (vgl. JENKINS 2006; JENSEN 2010; MEIKLE / YOUNG 2012). Medientheoretisch ließen sich diese Spannungen als verschiedene Intermedialitätsphasen deuten. Bei früheren Medienumbrüchen und -neuerungen lösten sich diese Phasen erst nach Jahrzehnten ab. Auch wenn die medientechnische Beschleunigung von immer leistungsfähigeren KI-Plattformen, Large Language Models (LLMs) und neuronalen Netzen beeindruckend ist, spricht doch derzeit nichts für eine ebensolche Beschleunigung von Intermedialitätsmomenten. Wie zu diskutieren sein wird, ist dies vielleicht auch kein Zufall, geht es bei dieser Ablöse doch gar nicht um technische Fragen (oder Innovationen), sondern um genuin kulturelle Momente der Integration neuer Medientechniken in bestehende und sich neu entwickelnde soziale ›Protokolle‹, Konventionen und Praxen, die eine viel höhere Halbwertszeit aufweisen als technische Updates. Aber natürlich ist die (Medien-) Geschichte des KI-Bildes noch nicht geschrieben und seine Medienarchäologie wird sicherlich viele verschüttete Irrwege freizulegen haben. Was also lässt sich, Stand Frühjahr 2024, zum Impact des KI-Bildes auf die Medienkonvergenz aussagen?

### 1.1 Primäre Intermedialität

Am Anfang soll man ja beginnen. Primäre Intermedialitätskonzepte treten, wie Rainer Leschke in seiner Typologie verschiedener Medientheoriegattungen argumentiert hat, »vom Objekt her motiviert« auf (LESCHKE 2003: 23), »quasi als heuristische Phase während der Latenz eines Mediums«; und zwar immer dann, wenn eine neuartige Technologie das Potenzial entfaltet, als Medienform gesellschaftlich wirksam und somit eine sozial, politisch oder kulturell relevante kommunikative Größe zu werden. Irina Rajewsky versteht unter diesem Moment fast gleichlautend ein »*kulturelles bzw. mediales Basisphänomen*« (RAJEWSKY 2008: 50; Herv. im Orig.). In dieser Phase erscheinen die kommunikativen, ästhetischen und sozialen Eigenschaften der neuen Technologie noch nicht entschieden und generell nur unzureichend erfasst. Abgrenzungen zu und Wechselwirkungen mit bestehenden Medientechniken scheinen daher noch im Flux. Aufgrund dieser ungesicherten ›Vermessungslage‹ erscheint die neue, eher noch potenzielle Medientechnologie *hybride*. Das »inter-« bzw. »zwischen-« dieses initialen Intermedialitätsmoments bezieht sich also auf den fraglichen ›Ort‹ und die fragliche Beschreibung des potenziellen Mediums zwischen bereits bestehenden Medienformen und deren konzeptuellen Vermessungen. Leschke sieht hier, in diesem vortheoretischen, zumeist essayistischen Stadium der Medientheoriebildung, häufig eine unsaubere Vermischung von Diskursen und Akteur*innen, insofern primäre Intermedialitätskonzepte zumeist nicht innerhalb von Universitäten, sondern an der Schnittstelle von Praxis und deren Reflexion vorgenommen werden – also etwa durch Praktiker*innen und Produzent*innen (vgl. LESCHKE 2003: 30). Von den typischen Momenten dieser Phase, die sich vom Auftreten der Fotografie über das Radio bis hin zum Computer immer wieder zu wiederholen scheint, sind folgende Aspekte besonders relevant für die Frage nach KI-Bildern:

Primäre Intermedialitätskonzepte verfahren zwangsläufig differenziell, indem relevante *Unterschiede zu Bestehendem* herausgearbeitet werden (vgl. dazu ausführlicher SCHRÖTER 2008: 590-594). »Auf die […] Einführung eines neuen Mediums wird mit vergleichenden Ansätzen reagiert, die den Kontrast zwischen den vorhandenen Medien und dem neuen, das den Anlass für die Reflexion bot, zu bestimmen suchen« (LESCHKE 2003: 23). Dieses Verfahren, so unumgänglich es in dieser Phase auch sein mag, ist damit auch höchst selektiv, bezieht ganz heterogene sinnliche wie technische oder soziale Eigenschaften mit ein und ist so wissenschaftlich eigentlich kaum gedeckt, gibt aber doch entscheidende Motive und Themen vor, die auch in anschließenden Phasen möglicher »Einzelmedienontologien« (LESCHKE 2003: 73-160) eine erstaunliche Beharrungskraft aufweisen können. »Was in den primären Intermedialitätsdiskursen festeschrieben wird, sind die Themen, die Motive und damit die Agenda der wissenschaftlichen Reflexion eines Mediums« (LESCHKE 2003: 70). Tatsächlich lässt sich die bereits angesprochene *IMAGE*-Aufsatzsammlung als exakt eine solche ›intermedial Vermessung‹ verstehen, um »die gröbsten Auffälligkeiten eines neuen

Mediums [durch] Serien von Differenzen und deren Bewertungen […] in den Griff zu bekommen« (LESCHKE 2003: 67). Die Autor*innen versuchen in verschiedenen Ansätzen und unter verschiedenen Perspektiven und Prämissen, das ›Neue‹ an KI-Bildern gegenüber bestehenden und bekannten Formen der Bildlichkeit zu konturieren.

Andreas Ervik etwa identifiziert sechs Dimensionen, unter denen KI-Bilder »radically distinct from other images« (ERVIK 2023: 53) erscheinen, welche Produktionsfaktoren (»Symbiogenesis: agency is shared between the prompting user, the platform holders, and the AI«, ERVIK 2023: 53, aufbauend auf einem »collective media imaginary« aus Datenbanken, ERVIK 2023: 54) mit formal-ästhetischen zusammenbringen wollen:

> *»Views of nowhere: Whereas the photographic presents a view of something from a singular point of view, the computer camera (of, for instance, a videogame) is untied from a unified, specific location and can instead display objects that can be rotated and potentially viewed from any angle. […] AI image generators offer something entirely different again. In contrast to either ›a view‹ or ›any view‹ of what is placed in front of a recording apparatus or produced with computer graphics, image generators could be said to produce multiple versions of views of nowhere« (Ervik 2023: 50).*

Dieses differenztheoretische Motiv wird auch aus anderen Perspektiven trianguliert, wenn etwa Roland Meyer eine veränderte Bedeutung von »style« diskutiert, welcher nicht mehr an kunsthistorische Epochen, Schulen, Individuen oder generell soziale bzw. historische Kontexte geknüpft sei, sondern zu einem »repeatable visual pattern extracted from the digitally mobilized images of the past« (MEYER 2023b: 100) geworden ist. »For these models, the ›photographic‹ seems to be just another ›style‹, an aesthetic, a certain ›look‹, not a privileged mode of indexical access to the world. And this ›photorealistic style‹, I would argue, simulates visual rather than optical aspects of the photographic« (MEYER 2023b: 108). Mit wieder anderen Worten gesprochen wird hier, als medien*spezifische* – und das bedeutet eben immer mediendifferenzielle – Eigenschaft von KI-Bildern eine besondere ›flatness‹ identifiziert, und zwar in doppelter Hinsicht: anders als frühere computergenerierte 3D-Simulationen mit anschließenden Bildausgaben kennt eine KI-Plattform *keinen* (auch nur virtuellen) *Bildraum*, sondern ausschließlich Pixel-*Oberflächen*.[1] Diese korrespondieren zugleich mit der metaphorischen ›Einebnung‹ aller sozialen Kontexte, Hierarchien und Diskurse. Eryk Salvaggio (2023b) beschrieb so auch einen bedenklichen medientechnischen *Kontext-Kollaps*, wenn Prompts wie »Archiv« in der LAION-Datenbank umstandslos und zugleich auf Fotos von Holocaust-Überlebenden wie von Wehrmachts-Soldaten und ebenso auf Episoden der Teenie-TV-Serie *Riverdale* zurückgreifen – einfach, weil all diese Bildartefakte irgendwann einmal mit einem entsprechenden Schlagwort versehen worden sind: »This is not a collage of images but a collage of documentation stripped of context, photographs without memory. It is stitching with cultural debris, pop culture and trauma woven into a single tapestry, the threading of the needle predicted pixel by pixel.« (SALVAGGIO 2023b: n.pag.). Wir sehen an diesen deutlich artikulierten *Bedenken* zugleich, dass in dieser primären Intermedialitätsphase auch grundsätzliche Bewertungen ausgesprochen und ausgelotet werden. Neben dem Kontext-Kollaps werden diese für KI-Bilder besonders in der technologischen Verstärkung von Vorurteilen und Klischees (vgl.

---

[1] Diese ›Physiklosigkeit‹ schreibt sich auch in den Bewegtbildern von Sora fort, die ebenfalls nur als Oberflächen wahrscheinlicher Pixel ohne jedes Tiefen- und Raumverständnis berechnet werden (vgl. MARCUS 2024).

SALVAGGIO 2023a),[2] im unentgoltenen Ausschöpfen menschlicher Arbeitskraft (vgl. MEYER 2023b) und in einer Verzerrung des kollektiven (und historischen) Bildgedächtnisses (vgl. OFFERT 2023) identifiziert.

Solche primären Intermedialitätsmomente lassen sich aber nicht nur, wie Leschke es tut, als Vorstufe oder Ausgangspunkt der medienwissenschaftlichen und medientheoretischen Reflexion verstehen, sondern noch grundlegender als Beschreibung von Praxen und Diskursen, die außerhalb der Wissenschaft aufzufinden wären. Tatsächlich wurde der Begriff ›Intermedia‹ bereits 1812 bei Samuel Taylor Coleridge verwendet und 1965/1981 durch den Künstler Dick Higgins so aufgegriffen, dass es dabei um eine Beschreibung *künstlerischer Praxen* gehen solle, »to define works which fall conceptually between media that are already known« (HIGGINS 2001: 52). Der Binnenmechanismus ist aber durchaus der gleiche wie später bei Leschke: eine neue ›Kunstform‹ – oder eher eine neue kommunikativ-technische Gattung, insofern es Higgins gerade auch um die Überschreitung von Kunstsphären, um eine Vermischungen mit anderen Diskurstypen wie Philosophie oder Alltagskommunikation geht – fällt gewissermaßen konzeptuell zwischen bestehende und bereits existierende soziale Praktiken. Dadurch wird sie zu diesem Zeitpunkt als hybride Provokation aufgefasst.

Higgins' kanonisierte Beispiele umfassen konkrete Poesie, Ready Mades, ›Mail Art‹ oder Fluxus-Performances. »Thus the happening developed as an intermedium, an uncharted land that lies collage, music and the theater. It is not governed by rules; each work determines its own medium and form according to its needs. The concept itself is better understood by what it is not, rather than what it is« (HIGGINS 2001: 50) Diese Differenzoperation ist stets historisch zu fassen, denn was in einem Jahr als »intermedial« erscheint, kann durch Wiederholung und Auskonventionalisierung wenig später etabliert sein. »[T]here is a tendency for intermedia to become media with familiarity. The visual novel is a pretty much recognizable form to us now. We have had many of them in the last 20 years« (HIGGINS 2001: 52). Deutliche Kriterien zur Nachvollziehbarkeit und Entscheidbarkeit blieb Higgins uns zwar schuldig, doch dürften sie auf diskursanalytischer Ebene zu verorten sein: In den Diskussionen um den adäquaten ›Ort‹ bzw. die angemessene Verwendung von KI-Bildern innerhalb und gegenüber bereits bestehenden medialen Bildformen. Für ›Diskursereignisse‹, die genau diesen problematischen, primär intermedialen Status von KI-Bildern innerhalb der weiteren Medienökologie deutlich machen, gibt es nun auch keinerlei Mangel. Geradezu für alle denkbaren (Bild-)Medienformen lassen sich Beispiele finden, in denen der Status der KI-Produktion eines Bildes zumindest dafür sorgte, dass von einer ›nahtloses‹ Integration – als etwa *eine* von unzähligen *anderen* Bildproduktionstechniken – zumindest derzeit keine Rede sein kann.

## 1.2 Indikatoren zur primären Intermedialität von KI-Bildern

Ein erstes, weithin diskutiertes und kommentiertes Diskursereignis war sicherlich die Kontroverse um die Auszeichnung des Bildes »Théâtre D'opéra Spatial« in der »Digital«-Kategorie des Kunstwettbewerbs der Colorado State Fair im September 2022. Prämiert wurde Jason M. Allen, der das Werk, ohne dies etwa zu verschleiern (»I didn't break any rules«, zit. nach ROOSE 2022: n.pag.), von Midjourney generieren ließ. Allens Autorschaft bzw. künstlerische Leistung beschränkte sich also auf das Experimentieren mit und Finden von geeigneten Prompts, die auch sorgsam geheim gehalten wurden sowie, noch grundlegender, auf die im Sommer 2022 noch einigermaßen innovativ Einarbeitung in Midjourney zum Zwecke der Erstellung eines solchen Bildes für diesen Kontext. Für Intermedialitätsfragen ist dabei weniger interessant, ob sich bei dem Resultat nun um ›Kunst oder Kitsch‹ handelt, sondern vielmehr, ob ein solcher *Bildtyp* mit generativer KI-Produktionsgeschichte für den Wettbewerb überhaupt zulässig ist (»Was it cheating?«, wie die *Washington Post* fragte, Harwell 2022: n.pag.). Zumindest dürfte wohl feststehen, dass »Théâtre D'opéra Spatial« *ohne* diese Kontroverse kaum die Aufmerksamkeit zahlloser internationaler Presseberichte in den größten Zeitungen der Welt erhalten hätte. Gleiches gilt wohl für den ähnlich gelagerten Fall von Boris Eldagsens »Pseudomnesia: The Electrician« (Abb. 1), Preisträgerbild beim Sony World Photography Awards 2023 in der Kategorie »creative open«, und ebenfalls in ›co-creation‹ mit einer unbenannten KI-Plattform erstellt. Wir haben also die Bezugskategorien der Bildtypen von ›digitalem Artwork‹ zu ›Fotografie‹ gewechselt, der Verlauf der Kontroverse ist auch noch etwas verzwickter (Eldagsen lehnte den Preis ab, da er die Jury angeblich nur »testen« wollte, die KI-Genese tatsächlich aber ebenfalls vorher enthüllt hatte; vgl. Glynn 2023). Muster und Resultat der Diskussion bleiben aber die gleichen: Egal, wie das Bildartefakt phänomenal aussieht, die Erstellung in ›Co-Autorschaft‹ mit einer KI-Plattform verschiebt das Resultat in einen unklaren, hybriden Zwischenstatus von Bildtypen oder Bildmedien, der eben noch entschieden werden muss, gerade aber durch diesen (gegenwärtigen) Entscheidungsdruck überhaupt erst internationale Aufmerksamkeit und Neuigkeitswerte generierte.

---

[2] Zum besorgniserregenden, offenbar derzeit technisch inhärenten Rassismus von Large Language Models, siehe auch HOFMAN ET AL. 2024.

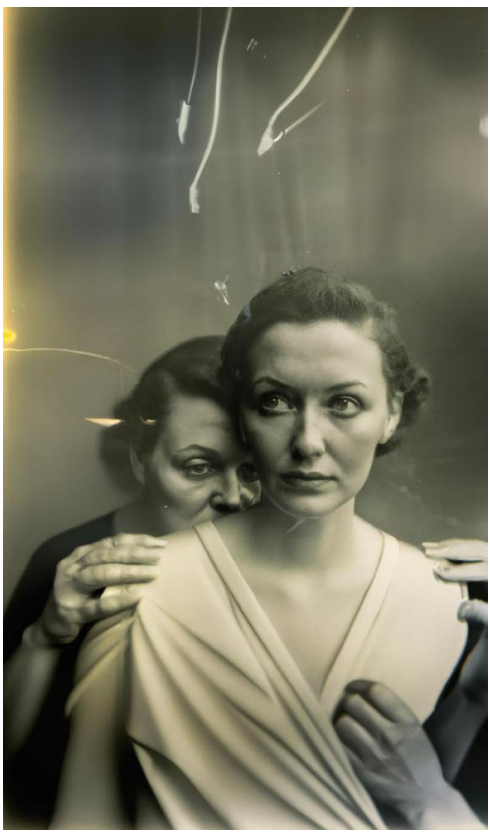

Abb. 1: Boris Eldagsens »Pseudomnesia: The Electrician« (vgl. Glynn 2023)

Nun sind Kunstwettbewerbe kaum repräsentativ für die weitere Medienökologie unter kommerziellem Produktionsdruck. Die angeführten Diskursmuster der primären Intermedialität bleiben aber auch überraschend stabil, wenn wir den bisherigen Einsatz von KI-Bildern in der Herstellung verschiedener populärer und konventioneller ›Einzelmedien‹-Produkte wie Comics, TV-Serien oder Videospielen betrachten – wenn auch vielleicht mit einigen überraschenden Akzenten. Als erstes KI-Comic gilt Kris Kashtanovas *Zarya of the Dawn*, das bereits recht früh (im Sommer 2022) exklusiv mit Midjourney-Bildern erstellt worden ist (vgl. KASHTANOVA 2022). Mit zeitlichen Abstand ist es wohl fair, zu urteilen, dass *Zarya of the Dawn*, als kleine Eigenproduktion, ohne diesen Spektakelwert kaum größere Beachtung gefunden hätte. An dieses Experiment haben sich seither zwar eine Reihe weiterer angeschlossen, zu nennen wären Carson Grubaughs *The Abolition of Man* (2022 im Verlag Living the Line) oder Tierry Murats *Initial_A* (2023 im Verlag Label Log Out). Die KI-Erstellung wird hier jeweils besonders ausgewiesen und als ›Verkaufsargument‹ beworben. Murats *Initial_A* eröffnet sogar mit einer großen, allenfalls leicht ironischen »Trigger Warning: AI Generated Images«. Für all diese Werke dürfte wohl gelten: Weder sind mir besonders wohlwollende Kritiken bekannt, die sich nicht auf den Neuigkeitswert der KI-Bilder (und stattdessen etwa auf deren narrative Qualitäten) fokussieren, noch wurde diese Produktionstechnik (bislang) von größeren Verlagen wie Marvel, DC oder Image aufgegriffen. Im Gegenteil lässt sich derzeit von einem öffentlichen Misstrauen sprechen, wenn mediale Mutmaßungen laut werden, eine Künstler*in *könnte* auf KI-Unterstützung zurückgegriffen haben. Im März 2024 etwa wurde auf *X/Twitter* darüber spekuliert (und zahlreiche forensische Indizienketten aufgebaut), ob Andrea Sorrentino für die Artworks von DCs *Batman*-Serie (Ausgabe #143) auf KI-Generatoren zurückgegriffen haben könnte, wo es um die Darstellungen von Figuren wie dem Joker ginge. Die Verantwortlichen haben dies weder bestätigt noch dementiert, der Verlag veröffentlichte aber zumindest ein Statement, dass dem nachgegangen werden solle: »[A]ll artwork must be the artist's original work. We are looking into the specifics of this situation« (zit. nach FLETCHER 2024: n.pag.). Müßig zu erwähnen, dass hybride Collagen unterschiedlichster Bild-Techniken (Handzeichnungen, Digitalzeichnungen, bearbeitete Fotos etc) im Comic im Kontrast ganz üblich sind und keinerlei Neuigkeits- und Spekulationswert besitzen.

Wesentlich transparenter sind einige KI-Einsatzversuche in Filmen und TV-Serien, hier lässt sich durchaus bereits von breiter angelegten Versuchen der Integration in die Medienkonvergenz sprechen. Im Netflix Japan/Wit Studio-Animationskurzfilm *The Dog and the Boy* vom Regisseur Ryotaro Makihara etwa wurden die gezeichneten Hintergründe KI-generiert – was im Februar 2023 umgehend für einen medialen ›outrage‹ gesorgt hatte: »Responses to the film were overwhelmingly negative, ranging from ›massive yikes‹ to ›shameful‹« (zit. nach CHEN 2023: n.pag.). Die bislang auffälligste Platzierung in der derzeit größten Produktion findet sich in der Marvel/Disney-Miniserie *Secret Invasion* (Juni/Juli 2023), Teil des »Marvel Cinematic Universe« und prominent besetzt mit bekannten Darsteller*innen wie Emilia Clarke und Samuel L. Jackson. Der zweiminütige, animierte Vorspann von Method Studios, der in derlei Serien generell ästhetisch klar vom Rest der Produktion abgegrenzt ist, basierte auf KI-Artworks. Auch wenn hier sicherlich vom Versuch einer nahtlosen Integration von KI-Bildern in der verteilten Produktion gesprochen werden muss, eint diese Beispiele dennoch abermals ein gewaltiger ›backlash‹: In den medialen Entrüstungen darüber, die keinen künstlerischen oder narrativen Mehrwert in diesen Integrationen erkennen lassen, äußert sich erneut eine fundamentale Kritik an der transparenten Strategie von Produktionsfirmen, lediglich Kosten einzusparen wollen und menschliche Zeichner*innen bzw. Animator*innen um Lohn und Gehalt zu bringen. »›I'm devastated,‹ tweeted Jeff Simpson, a visual development concept artist who said he worked on the character design and props for ›Secret Invasion,‹ though not the intro sequence. ›I believe AI to be unethical, dangerous and designed solely to eliminate artists' careers.‹« (zit. nach SCRIBNER 2023: n.pag.).

Überraschen mag es, dass dieser Befund nicht einmal sonderlich anders ausfällt, wenn man in den Bereich der Videospiele-Produktion blickt, wo man vielleicht eine höhere Affinität zu ›technischen Bildern‹ und geringere Erwartungen an handwerklich-künstlerische Leistungen erwarten würde. Doch auch die Produktionsfirma Evercade sah sich im Juni 2023 genötigt, den Trailer und die Artworks zu einem *Duke Nukem*-Re-Release zurückzuziehen und sich öffentlich dafür zu entschuldigen, nachdem Fans bemerkt hatten, dass der verantwortliche »hybrid artist« Oskar Manuel für die Erstellung der Bilder auf KI-Plattformen zurückgegriffen hatte (vgl. PLUNKETT 2023). Im Januar 2024 war sogar von einer »AI Hysteria« die Rede, nachdem dem japanischen Spiel *Palword* von Pocket Fair (offenbar fälschlich) unterstellt wurde, für seine character designs auf KI-Generatoren zurückgegriffen zu haben. Nachdem Fans bereits zum Boykott des Spiels aufriefen, markierte der Spiele-Journalist Ryan Broderick die Ironie der Vorwürfe für diesen Medienbereich: »It would honestly take more effort to use generative AI to make 3D video game character designs than to just pay someone to make bad ones« (BRODERICK 2024: n.pag.). Wie dem auch sein mag: an dem übergreifenden Befund der derzeitigen Diskursmuster scheint sich abermals auch hier nichts zu ändern. Der geeignete ›Ort‹ für KI-Bilder innerhalb der Medienökologie ist trotz allen Konvergenzbestrebungen noch nicht ausgemacht. KI-Bilder markieren stets, im besten Fall, einen spektakulären Ausnahmefall, im schlechtesten (und wesentlich häufigeren) Fall eine Kontroverse und einen medialen Backlash, in beinahe keinem bislang bekannten Fall einen ästhetisch oder narrativen Mehrwert, der als solcher normalisiert und konventionalisiert werden könnte.

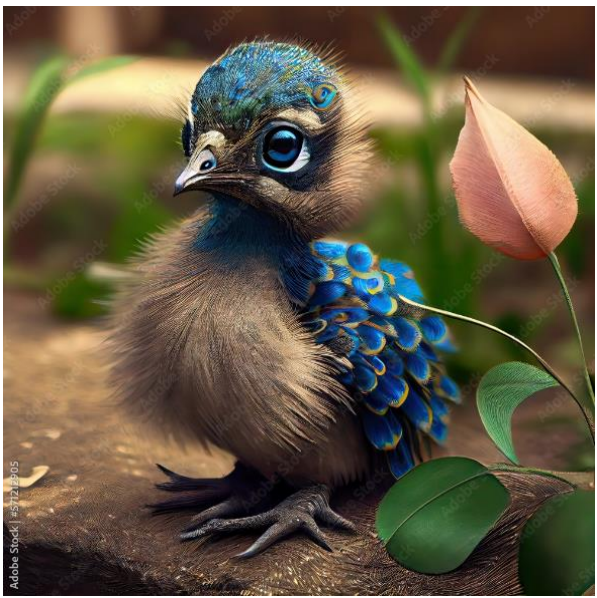

Abb. 2: Der berüchtigte »Baby Peacock«, der in Wirklichkeit natürlich nicht so aussieht (vgl. LAMAGDELEINE 2023)

Als letzten Indikator für eine solche primäre Intermedialität von KI-Bildern zwischen bestehenden Medienformen und Dispositiven könnte man auch die großen Irritationen heranziehen, welche KI-Bilder (immer häufiger) inmitten von Google-Bildersuchergebnissen auslösen. Viel diskutierte Beispiele sind die Ergebnisse nach Suchen zum hawaiianischen Sänger und Ukulele-Spieler Israel Kamakawiwoʻole, die für viele Nutzer*innen an erster Stelle bizarre Midjourney-Imaginationen aufspülen (vgl. Larsen 2023); Bilder vom »Baby Peacock«, die eben kein tatsächliches Exemplar dieser Gattung, sondern seit April 2023 immer häufiger und ohne weitere Kennzeichnung eine Art fiktives ›Pokemon‹ ausfindig machen, in dem das Aussehen eines erwachsenen, männlichen Pfaus mit den Attributen von »cuteness« verschmolzen wurden (vgl. LaMagdeleine 2023, Abb. 2); und die Kontamination unseres historischen Gedächtnisses, wenn Suchen nach geschichtlichen Ereignissen wie dem Tianaman Square-Massaker zu KI-erzeugten ›Selfies‹ statt zu dem ikonischen Dokumentarfoto von Jeff Widener führen (vgl. Growcoot 2023). Die Sorgen hinter den entsprechenden Zeitungs- und Onlineartikeln, die eine Verunreinigung unseres kulturellen Bildgedächtnisses, eine haltlose Vermischung von Fakt und Fiktion und letztlich ein Abdriften in die Welt der Simulakren fürchten, sind verständlich. Umgekehrt aber scheint dem eine eigentümliche These bzw. eine Erwartung der generellen Non-Fiktionalität von Google-Bildersuchen zu unterliegen, welche durch die Intervention von KI-Bildern überhaupt erst thematisch wird. Annahmen an Non-Fiktionalität von generellen Bildersuchmaschinen sind schließlich weder theoretisch zu plausibilisieren noch von Anbieterseite versprochen oder technisch begründbar. Auch hier scheinen KI-Bilder seltsam zu intervenieren und gewohnte Medienbereiche und ihre Verwendungsweisen (›Protokolle‹) durcheinander zu bringen, ohne dass sich eine baldige Stabilisierung abzeichnen lässt. Die Rufe nach einer generellen Kennzeichnungspflicht von KI-Bildern, so nachvollziehbar sie in Zeiten von uferloser Desinformation und visuellen Propagandamöglichkeiten auch sein mögen, scheinen so zunächst einem generellen Bedürfnis Ausdruck zu verleihen, KI-Bilder *als Medienformen* zu identifizieren und zu unterscheiden, um überhaupt erst damit beginnen zu können, ihnen einen geeigneten Ort innerhalb unserer Mediendispositive zuzuweisen – und sie *dann* gegebenenfalls mit anderen, »konventionell als distinkt wahrgenommene[n] Kunst- oder Mediengattungen« (Rajewski 2008: 55) oder »qualifizierten Medien« (Ellestöm 2010: 25; Übers. L.W.) interagieren zu lassen, ohne dass dies die derzeit noch allgegenwärtigen Kontroversen auslöst.

### 1.3 Exkurs: Multimodale KI-Bilderzeugung und Medientheorie

Man mag versucht sein, diese Hybridität und ›mediale Widerständigkeit‹ von KI-Bildern in der multimodalen Technologie von text2image-Plattformen begründet zu sehen, also in der undurchsichtigen Blackbox, in der verbale Prompts in einem ersten Schritt von Text-Encodern mit Large Language Models (wie CLIP) verarbeitet, in einem zweiten Schritt zu verschieden trainierten Diffusion-Modellen (wie CivitAI, Juggernaut SL, MeinaMix oder Deliberate bei Stable Diffusion) weitergereicht und in Schritt drei von Image Decodern ›in Pixeln gemalt‹ werden (vgl. etwa für Stable Diffusion die hilfreiche Übersichtsdarstellung von Alammar 2022; allgemeiner Bajohr 2024); diese drei unterschiedlichen Prozesse wiederum bauen alle auf Datenbanken auf (wie LAION), die selbst bereits multimodal sind, also aus Text/Image-Paaren bestehen, mit denen sowohl Text-Modelle als auch Bild-Modelle trainiert werden können. Was in Dall·E wie eine einzelne Blackbox mit festem Input (Prompts) und Output (KI-Bilder) erscheint, lässt sich in open source-Varianten wie Stable Diffusion in allen Einzelelementen modifizieren und variieren.

Für die gegenwärtige Anmutung primärer Intermedialität und Hybridität ist dies aber eher unerheblich. Wie Higgins bereits 1965 anmerkte, geht es dabei um einen ganz und gar kulturellen (und eben nicht technischen), wahrgenommen ›Einheits-Makel‹. Es geht um eine *konzeptuelle* Einheit von Produktionstechniken, Materialien, Institutionen und sozialen Funktionen, »[t]he idea that a painting is made of paint on canvas or that a sculpture should not be painted« (HIGGINS 2001: 49). Die faktische Multimodalität spielt dafür kaum eine Rolle, Higgins grenzt daher »intermedia« auch scharf von (ebenso multimodalen) »mixed media« wie der Oper ab: »Many fine works are being done in mixed media (…). But one knows which is which« (HIGGINS 2001: 52). Wenn ein jedes ›Einzelmedium‹, wie medienwissenschaftlich ganz unstrittig sein sollte, allenfalls *konventionell als distinkt verstanden* werden kann, sich bei genauerer Betrachtung jedoch stets als hybride Konfiguration, als Dispositiv oder Assemblage von heterogenen Elementen erweist, so wären Ansätze zur medientheoretischen Konzeption von KI-Bildern ebenfalls in Akteur-Netzwerk-Perspektiven oder Akteur-Medien-Perspektiven zu suchen (vgl. SCHÜTTPELZ 2013; KRIEGER / BELLIGER 2014; SPÖHRER / OCHSNER 2017). Überlegungen dazu sind durchaus bereits vorhanden (siehe WILDE 2023; BOLTER 2023). Aber auch die Probleme mit derlei Konzeptionen sind hinreichend diskutiert (vgl. JUNG / SACHS-HOMBACH / WILDE 2021). Jedes ›Netzwerk‹ der Akteur-*Netzwerk*-Theorie ist prinzipiell unendlich, es existiert eben nur *ein* Netzwerk – *die Welt*. Die verteilte Autor*innschaft von KI-Bildern müsste damit, wie Hannes Bajohr (2023b) gezeigt hat, ebenfalls in uferlos unterschiedlichen Größen gesucht werden, die von den ›promptenden‹ Nutzer*innen über das training set und deren millionenfachen Urheber*innen über die Programmierer*innen der Modelle bis zu Unternehmen wie OpenAI reichen. Nicht nur das:

> »Finally, the network would even include all the matter involved: the machines that run the code as well as the minerals and rare earths they are made of, which are extracted from the soil in a process that emits CO2, a direct factor in climate change. From this vantage point, the Earth itself would lay claim to be a participant in the authorship network« (BAJOHR 2023b: 20-21).

Eingrenzungen lassen sich somit allenfalls forschungspragmatisch vornehmen, etwa durch eine grundsätzlich zwar arbiträre, mindestens aber ethisch gebotene Ergänzung von ANT-Modellen durch (doch wieder) anthropozentrische Autor*innenschaftskonzeptionen. Erneut aber, und das ist der springende Punkt, ist dies keinesfalls ein Spezifikum von KI-Bildern, sondern träfe auf jedes beliebige ›Einzelmedium‹ zu.

Verkürzt gesprochen: Nicht erst KI-Bilder sind ›eigentlich‹ (durch ein angenommenes neues technologisches Primat) hybride; selbst eine vergleichsweise handfest-materielle Medienform wie ›der Comic‹ zerfällt bei näherer Betrachtung in eine Assemblage verteilter Elemente, Akteure (und Akteur*innen) und Handlungsächte (vgl. OSSA / THON / WILDE 2022). Dass ›Film‹ natürlich auch alles andere als monolithisch ist und im Produktions- wie auch Distributions- und Rezeptionsprozess bestens unter Akteur-Netzwerk-Perspektiven untersucht werden kann – und auch ästhetisch in Tonspur und Bildspur, zahlreiche kombinierten Zeichensysteme wie Bilder, Sprache und genuin filmischen Gestaltungsmittel ›zerfällt‹ – ist beinahe schon trivial (vgl. SPÖHRER 2017). Ebenso aber: Auch wenn Comic und Film heute beide im gleichen medialen ›Aggregatszustand‹ (nämlich digitalen Datensätzen) distribuiert und im gleichen technischen Apparat (etwa dem Smartphone) rezipiert werden (können), unterscheiden wir beide doch weiterhin als grundsätzlich verschiedene Medienformen. Sie werden typischerweise sehr unterschiedlich produziert und sind an verschiedene sozialsystemische Institutionen angeschlossen. Es existieren zudem auch medienästhetisch betonte, historisch auskonventionalisierte und häufig gezielt inszenierte Differenzen, an die divergierende Praktiken und Protokolle sowie kulturelle Valenzen gekoppelt bleiben (vgl. LESCHKE 2014).

Wenn sich die Frage nach Intermedialitätsphasen also nicht technologisch vorentscheiden lässt, so verschiebt sie sich grundsätzlich in Richtung einer soziokulturellen »Phänomenologie und Alltagshermeneutik der Situation« (MÜLLER 1996: 105). ›Einzelmedienkonfigurationen‹ (konventionell als distinkt wahrgenommene Kunst- oder Mediengattungen bzw. qualifizierte Medien) sind dann nichts anderes als ›frames‹, durch die eine institutionalisierte Einheit von Produktion, Technik und Setting hergestellt und irgendwann vorausgesetzt werden kann (vgl. DETERDING 2013). Jedes einzelne mediale Artefakt oder jede einzelne mediierte Interaktion ließe sich demnach stets anhand verschiedener, grundsätzlich disparater Medialitätsebenen erfassen und beschreiben (etwa semiotisch-kommunikative Medialität, technisch-apparative Medialität und sozial-institutionelle Medialität, vgl. THON 2014; JUNG / SACHS-HOMBACH / WILDE 2021), von denen stets nur bestimmte, und eben zwangsläufig selektive und kontingente Momente relevant zur Auszeichnung *dieser* Medialität in gegebenen Kontexten scheinen. Einzelmedien müssen demnach immer

> *»diskursiv aus einem heterogenen Netzwerk […] von technischen Verfahren, Institutionen, Programmen, Diskursen […], formale[n] Strategien, Autorenfiguren, Praktiken etc. je nach einem bestimmten strategischen Zweck ›herausgeschnitten‹ [werden]« (SCHRÖTER 2008: 594).*

Die dafür relevanten ›Schablonen‹ sind häufig zumeist die Elemente, die im primären Intermedialitätsdiskurs bereits als Motive festgeschrieben worden sind. Und eben anhand dieser ›Marker‹ von relevanten Unterschieden können nun wiederum Mediengrenzen ästhetisch-kommunikativ, technisch-apparativ oder sozial-institutionell ausgestellt, ausgehandelt oder gar (re-)inszeniert werden. Hier also nun wären wir im Bereich der sekundären Intermedialität angekommen, die – so die Ausgangsthese – im Bereich KI-Bildlichkeit noch nicht erschlossen ist, durchaus aber bereits anzudenken wäre.

### 1.4 Sekundäre Intermedialität

Während in der Phase primärer Intermedialitätsdiskurse praxis- und produktionsnahe Beschreibung (hinzuzufügen wäre hier sicherlich auch ›marketing speech‹) und wissenschaftliche Reflexion zwangsläufig nicht sauber voneinander zu trennen sind, erscheint dies in der später daran anschließenden sekundären Intermedialität deutlich leichter möglich. Hier kann nun einerseits vorausgesetzt werden, dass die Grenzen von Medienformen und Medienbereichen bereits gezogen und einigermaßen unstrittig sind, dass andererseits so auch ästhetische Verfahren, mediendiskursive Aushandlungen und wissenschaftliche Reflexionen (als Intermedialitätstheorien) deutlich voneinander zu unterscheiden wären.

> *»Primäre Intermedialität hat es also auf das Ziehen von Grenzen und die Reglementierung eines neuen Mediums abgesehen und nicht auf die Erweiterung von Grenzen. In der sekundären Intermedialität hat sich demgegenüber jedoch der Fokus geändert; es geht nicht mehr um Differenzen und die unterschiedliche Bedeutsamkeit von Medien, sondern um das Gegenteil, um Indifferenz und Interferenzen«* (LESCHKE 2003: 309)

Erneut wäre zu beachten, dass dies keine technische bzw. technologische Frage ist. »Das neue intermediale Terrain, das sich nicht mehr an den Rändern der Einzelmedien, sondern zwischen ihnen befindet, hat sich von der Materialität und Technizität der jeweiligen Bezugsmedien weitgehend gelöst« (LESCHKE 2003: 310). Mag sein, dass das ›mediale Substrat‹ sowohl von semiotischen Modalitäten wie Bildern, Texten oder Musik als auch von ganzen konventionalisierten Medienformen wie Filmen, Comics oder Hörbüchern auf der gleichen digitalen Verfasstheit besteht und sich in der ›universellen Maschine‹ des Computers nur noch als wechselnde Interface-Effekte zeigt. Hatte die generelle Medientheorie des Computers bzw. der Digitalität genau darauf beharrt und insistiert, dass Einzelmediengrenzen eigentlich obsolet geworden seien, bemühen sich sekundäre Intermedialitätstheorien nun in Reaktion, »neuerlich den Differenzen zwischen den Medien nachzuspüren« (LESCHKE 2003: 28) – weil diese konventionellen Mediengrenzen in der Praxis eben weiterhin relevant sind und sich sowohl diskursanalytisch als auch ästhetisch sehr gut nachzeichnen lassen.

Es ginge im Falle von KI-Bildern also zunächst einmal um ihre Identifizierbarkeit qua unterschiedlichen Markierungsverfahren *als KI-Bilder* in verschiedenen medialen Kontexten. Im umgekehrten Fall der medialen Täuschung – wenn sie als Fotografien oder als Handzeichnungen ausgegeben und missverstanden werden – entgeht uns schließlich genau ihre differenzbildende Qualität. Ebenso, wie Bilder gerade nicht *als Bilder* verstanden werden, wenn wir ihnen im Moment der Zeuxis-Täuschung auf dem Leim gehen, ebenso müssen wir KI-Bilder als solche erkennen.

Daran anschließen ließe sich die bekannte Trias der (sekundären) Intermedialitätsbeziehungen von Rajewsky (2008: 53-55): In »Medienkombination[en]« (was wir mit Higgins bereits als »mixed media« diskutiert hatten) würden KI-Bilder zwar ihre Erkennbarkeit erhalten, aber als unproblematischer Bestandteil anderer Medienformen einen ästhetischen (oder narrativen oder kommunikativen) Mehrwert dadurch generieren, dass es sich um KI-erzeugte Ansichten handelt. Schröter (2008: 130) spricht hierbei von »synthetischer« Intermedialität. Wäre der KI-Vorspann von *Secret Wars* etwa begeistert als solcher aufgenommen und als ästhetische Neuerung akzeptiert werden – mit daran anschließenden Traditionsbildungen – so könnte davon vielleicht die Rede sein. Dies ist ersichtlich noch keineswegs der Fall, und auch das Genre der KI-Comics stellt derzeit allenfalls ein Kuriosum dar. Rajewskys zweite Kategorie des Medienwechsels und der geg. daran anschließbaren Transmedialität (mit SCHRÖTER 2008: 136 »formale Intermedialität«) scheint noch weiter entfernt, meint diese doch eine Zirkulationsbewegung von Inhalten, Stoffen, Formen oder Motiven über als distinkt wahrgenommene und unstrittig vorausgesetzte Mediengrenzen hinweg. Als ein solches ›distinktes Mediendispositiv‹ kann bei KI-Bildern noch nicht gesprochen werden, eben weil die Koppelung von Technik, Praxen, Rezeption, Setting und schließlich Genres und Gattungen sich noch nicht einmal abzeichnet. Am interessantesten scheint mir derzeit die Kategorie der (impliziten) intermedialen Bezugnahmen (»transformationale Intermedialität«, SCHRÖTER 2008: 144), wo also distinkt wahrgenommene Mediendispositive sich gegenseitig referenzialisieren. »Hier könnte man … von ›Re-Representation‹ […] sprechen: Die intermediale Beziehung besteht dann darin, daß ein Medium ein anderes repräsentiert« (SCHRÖTER 2008: 144).

Bei KI-Bildern scheint dies durchaus eine inhärente medienästhetische Dynamik zu sein, insofern diese doch sehr häufig nicht nur einen im Prompt referenzialisierten ›Content‹, sondern auch einen bestimmten ›Style‹ remediieren. Roland Meyer (2023b) argumentierte, wie bereits angeführt wurde, dass ein neues, radikal entkontextualisiertes Verständnis von ›Style‹ zur eigentlichen Ware von text-to-image-Generatoren geworden sei.

> *»[F]or these models, everything, individual artistic modes of expression, the visual stereotypes of commercial genres, as well as the specific look of older technical media like film or photography, becomes a recognizable and marketable ›style‹, a repeatable visual pattern extracted from the digitally mobilized images of the past« (MEYER 2023b: 100).*

Die heute bereits viel studierte Ressource *The DALL·E 2 Prompt Book* vom Juli 2022 (DALL·ERY GALL·ERY 2022), ein frühes Tutorial zur kreativen Verwendung der damals neuen Plattform, präsentiert im Hauptteil bekanntlich eine Sammlung von Stichwörtern mit Beispielkreationen zu hunderten solcher ›remediierter‹ Styles. Ebenso viel diskutiert ist sicherlich, dass der Name des polnischen Fantasy-Illustrators Greg Rutkowsky (bekannt für u.a. Dungeons & Dragons und Magic: The Gathering-Artworks) lange Zeit einer der am meisten verwendeten Midjourney-Prompts war (selbstredend, ohne dass der Künstler dafür je rekompensiert worden wäre oder um Erlaubnis gefragt wurde; vgl. HEIKKILÄ 2022). Daher ist es auch nur folgerichtig, wenn Meyer selbst die scheinbar remediationsfreien ›transparenten‹ (fotografisch anmutenden) KI-Bilder als bloßen Sonderfall einordnet: »For these models, the ›photographic‹ seems to be just another ›style‹, an aesthetic, a certain ›look‹, not a privileged mode of indexical access to the world« (MEYER 2023b: 108).[3]

### 1.5 Indikatoren zur sekundären Intermedialität von KI-Bildern

Diese Remediationen scheinen aber bislang nur in eine Richtung zu funktionieren – oder zumindest nur in ganz wenigen Ansätzen ›umgekehrt‹ herum, als intermediale Bezugnahmen *auf* KI-Bilder im Paradigma sekundärer Intermedialität! Denn damit würde sich erneut und umso stärker die Frage nach der konkreten Erkennbarkeit und Identifizierbarkeit von KI-Bildlichkeit stellen, etwa durch ästhetische Verfahren (vgl. RAJEWSKY 2008: 57).[4] Wodurch wissen wir – wenn nicht anhand der expliziten

---

[3] In Bolter und Grusins (2002) bekannter Terminologie hätten wir es hier mit der anderen Seite der Remediationsrhetorik zu tun, die auf *immediacy* (scheinbare Unmittelbarkeit oder Nicht-Mediatisiertheit, BOLTER / GRUSIN 2002: 318) statt auf *hypermediacy* (ein gezieltes Ausstellen der medialen Darstellungsbedingungen, BOLTER / GRUSIN 2002: 335) abzielt – beides freilich innerhalb der Prämissen der Remediation.

[4] Interessanterweise haben wir nun die Bezugsebene gewechselt und schauen, mit Rajewsky (2008: 48) gesprochen, nicht mehr auf »Fragen der Mediengenealogie, der Medienerkenntnis oder der grundsätzlichen Funktionslogik von Medien«, sondern auf Strategien und Bezüge innerhalb konkreter Texte, Bilder bzw. ›Werke‹: »Intermedialität als Kategorie für die konkrete Analyse medialer Konfigurationen« (RAJEWSKY 2008: 49). Sekundäre Intermedialität ist daher bei ihr auch konsequent eine Praxis »literatur- und kunstwissenschaftlicher Provenienz« (RAJEWSKY 2008: 48) und Leschke (2003: 299-326) ordnete sie bereits den »Medienphilologien« bei.

Thematisierung durch Sprache – dass ein anderes Bildmedium KI-Bilder thematisiert, imitiert oder kritisiert? Diese Erkennbarkeit würde sich gewiss abermals auf allen drei Ebenen von Medialität untersuchen lassen, ästhetisch-kommunikativ, technisch-apparativ sowie sozial-institutionell. Die folgenden Beispiele sollen dabei nicht den Eindruck von Vollständigkeit der Verfahren erwecken (die, soweit sie künstlerischer Natur sind, sicherlich ohnehin kaum Regeln unterliegen werden, insofern sie nur im Kontext eines spezifischen Artefakts ›funktionieren‹ müssen). Sie können aber zumindest andeuten, wie eine solche sekundäre Intermedialität, in der KI-Bildlichkeit einen festen Platz zugewiesen hätte, grundsätzlich aussehen könnte.

Der ›Königsweg der Medienreflexion‹ ließe sich einmal mehr im ›Störungs‹-Paradigma ausmachen (vgl. KÜMMEL / SCHÜTTPELZ 2003), also einem Zusammenbrechen ansonsten transparenter Medialität. Bei digitalen Bildmedien werden darunter oft Bugs und Glitches verstanden, die zunächst unintendiert auftreten und dabei die verborgene technisch-apparative Genese eines Bildes unfreiwillig ›enthüllen‹ können. Das typischste KI-Bildproblem wurde lange Zeit beispielsweise in einer falschen Anzahl von Fingern gesehen, womit KI-Bilder zumindest bis zum Frühjahr 2023 noch große Schwierigkeiten hatten. Amanda Wasiliewsky (2023) arbeitete die technischen Gründe für diese Glitch-Anfälligkeit heraus. Dies ist freilich lediglich eine Frage der jeweiligen technischen Iteration, bereits Midjourney 5.2 (veröffentlicht im Juni 2023) hatte deutlich seltener Probleme mit der korrekten Darstellung von Fingern. Dies freilich befeuert nur die Notwendigkeit der immer genaueren Suche nach immer neuen Indizien, die unintendiert auf KI-Medialität schließen lassen. »Spotting A.I. 📷s is an important media literacy skill. […] Let's learn how!«-threads (CHIYKOWSKI 2024) und ähnliche Artikel werden uns daher sicher noch lange begleiten.

Kulturell können derartige Auffälligkeiten aber deutlich länger ›nachwirken‹, so dass eine falsche Fingerzahl sicherlich noch für einige Zeit als als ›Markierung‹ von KI-Genese gelesen werden kann. Und als solche kann sie nun eben auch in anderen (gerade *nicht* KI-generierten Bildmedien) imitiert werden. Ein humorvolles Beispiel dafür findet sich in einem weit geteilten Facebook-, Twitter/X- und Reddit-Meme aus dem Februar 2023, in dem ein ›prosthetischer Finger-Ring‹ präsentiert wird, mit dessen Hilfe Träger*innen fotografisches Material juristisch unbrauchbar machen könnten, da jedes (tatsächliche, indexikalische) Foto durch den sechsten Finger für ein KI-Bild missverstanden werde (vgl. DAN 2023). In der humorvoll imaginierten, fiktiven Situation, in der »criminals« sich so der fotografischen Dokumentation entziehen, liegt keine erkennbare, ausgewiesene intermediale Bezugnahme (sondern eben eine Täuschung) vor. Das Meme reflektiert aber für seine schmunzelnden Betrachter*innen die veränderten medialen Bedingungen, unter denen bestimmte fotografisch festgehaltene Bildaspekte nun auf KI-Medialität schließen lassen – eben im Rückschluss auf unintendierte Bugs und Glitches, die zu diesem Zeitpunkt bereits als konventionalisiert gelten können.

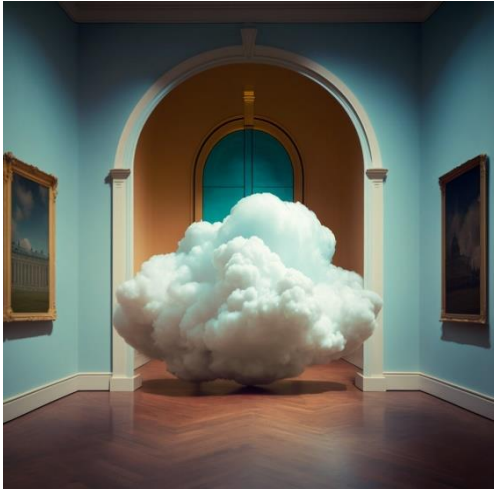

Abb. 3: »fluffy glamour glow« (MEYER 2023a)

Auf ästhetischer Ebene ist die Frage nach KI-Bildlichkeit zunächst noch schwieriger zu beantworten, wenn eine Remediation (und Imitation) unterschiedlicher ›Styles‹ doch ein ganz grundlegender Aspekt der KI-Bildgeneration darstellt. Wie also sollte von ›einer‹ KI-Ästhetik die Rede sein, die erkannt und gegebenenfalls auch imitiert (intermedial referenzialisiert) werden könnte? Und doch gibt es deutliche Anzeichen für eine Konventionalisierung gerade auch auf dieser Ebene. Denn selbst wenn KI-Bilder *theoretisch* jeden beliebigen ›Style‹ imitieren können, sorgen Nutzungs-Präferenzen, kommerzielle Zielsetzungen und Datenbank-Zusammensetzung doch für deutliche Musterbildungen. Meyer sprach in Rückgriff auf den Künstler Nils Pooker (2023) von einem »fluffy glamour glow« mit einer erkennbaren Farbpalette (»Teal and Orange«), durch die sich KI-Bilder häufig auszeichnen (MEYER 2023a, Abb. 3). Jede Plattform kann natürlich (in jeder Version bzw. jedem Modell) andere Schwerpunkte setzen und auch hier lassen sich wieder Wechselwirkungen und Verschiebungen ausmachen.[5] Dennoch haben sich unweigerlich die Koordinaten verschoben. Bestimmte Bildästhetiken werden nun als ›typisch KI‹ angesehen, insbesondere, wenn sie auf Mobilgeräteanzeige optimierte Kontrastfarben aufweisen. Aufschlussreich ist hier etwa die Debatte um die Superbowl 2024-Fotoserie »He Gets Us« (vgl. POLITIFACT 2024). Die Fotografin Julia Fullerton-Batten wies auf ihren Social Media-Kanälen nach, dass die Bilder tatsächlich in sehr teuren und aufwändigen Fotoshootings entstanden waren, die »highly cinematic visual storytelling«-Ästhetik (POLITIFACT 2024: n.pag.) zu diesem Zeitpunkt allerdings als (hier: fälschlicher) Marker für KI-Erzeugung gelesen wird.[6] Derlei ästhetische Konventionalisierungen werden also sicherlich weiterhin für unterschiedliche intermediale Lesbarkeiten von ›KI vs. Non-KI-Bildern‹ sorgen.

Auf sozial-institutioneller Ebene schließlich fragen wir nach der der Referenzialisierbarkeit sich etablierender ›Protokolle‹ der KI-Bildproduktion und -Distribution (zwischen spezifisch verteilten menschlichen und institutionellen Akteur*innen), also »[n]orms about how and where one uses [a media form … ], normative rules and default conditions« (GITELMAN 2008: 5, 7). Eine Thematisierung und Reflexion der konventionalisierten Verwendungsweisen von KI-Bildern wäre wohl vor allem auf Ebene von Narration zu erwarten, wenn also erzählende Medientexte wie Romane, Filme, TV-Serien oder Comics die mediale Gegenwart eingeholt haben und uns wie selbstverständlich vom Umgang mit ›prompt engineering‹ und KI-Plattformen berichten – und dies vielleicht sogar bildlich reflektieren. Comic Strips, Editorial Cartoons und Memes bilden hier wie üblich die Speerspitze der narrativen Kommentierung. Ein pointiertes Beispiel für eine solche Medienreflexion wäre ein Meme, das im Dezember 2023 ohne Angabe einer Urheber*in in der Facebookgruppe »AI Revolution - MidJourney AI, Dall·E 2, Stable Diffusion« gepostet wurde und das zwei Höhlenmenschen vor einer Felsenmalerei zeigt (vgl. STARK 2023). Der Bildkommentar liest sich: »That's not Art. You just copied the Bison«. Parodiert und kritisiert wird hier offenbar der Vorwurf, KI-Bildlichkeit könne nicht kunstförmig sein, da es sich nicht um ›Original-Schöpfungen‹ handele, indem der gleiche Vorwurf an der Urszene der allerersten Menschheitsbilder ad absurdum geführt werden soll. Vor dem vorgeschlagenen Tertium Comparationis (eines allgemeinen Weltbezugs in der Bildproduktion) sollen somit alle medientechnischen, ästhetischen und sozialen Unterschiede zwischen Handzeichnung und KI-Genese eingeebnet werden. Ein vergleichsweise komplexeres Beispiel aus der Social Media-Kommunikation – mit umgekehrter Stoßrichtung, nämlich einer Kritik von gegenwärtigen KI-Bildnutzungen – wäre ein 1 Million mal aufgerufener Tweet aus dem Frühjahr 2024 (SOUTHEN 2024, Abb. 4), in welchem ein Screenshot zum angemahnten ›Plagiieren von Bild-Prompts‹ kommentiert wird, welcher warnte: »Before you steal AI. generated images realize that: a prompt was not created within 10 seconds«. Der bissige Kommentar »This was posted unironically, what a world 😂« kontrastiert die fragliche Schöpfungshöhe von KI-Bildern via Prompts mit dem unhinterfragten Imitieren von Bildmustern in den Datenbanken von OpenAI und Co., ohne dass je Einverständnis, Kompensation oder bloß Kenntnis der ursprünglichen Schöpfer*innen vorlegen hatte. Ein Wissen um den gesamten medialen, institutionellen und wirtschaftlichen Kontext der Produktion, Distribution und Rezeption von KI-Bildern wird in beiden Beispielen bereits vorausgesetzt und zum Material der Medienreflexion gemacht.

---

[5] Meyer sah etwa eine »midjourneyfication« (2023a: n.pag.) in einem Dall·E Update vom März 2023 (während Dall·E3 im Oktober des Jahres wieder deutlicher in Richtung »Stock Illustration« tendierte).
[6] Diesen Punkt arbeitete Roland Meyer in seinem Online-Gastvortrag »Platform Realism: AI Image Synthesis and the Rise of Generic Visual Content« im NTNU »AI Media« Research Network am 16. Februar 2024 heraus.

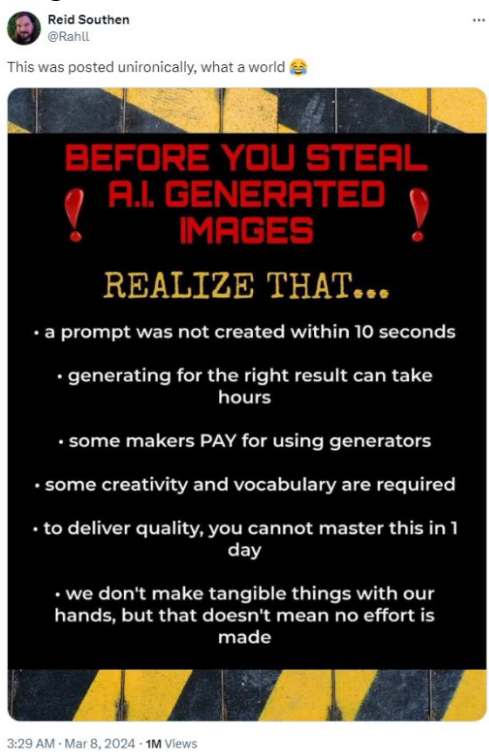

Abb. 4: »This was posted unironically, what a world 😂« (SOUTHEN 2024)

Damit die Protokolle der KI-Bildverwendung tatsächlich thematisiert und intermedial referenzialisiert werden könnten, müsste freilich zunächst mal ihr kultureller Einsatz hinreichend klar sein – und wie im vorigen Abschnitt diskutiert worden ist, ist diese Aushandlung derzeit noch in vielen Bereichen offen. Wir wissen, wie KI-Bilder erstellt werden, aber kaum, zu welchem Zweck, denn ein ›unproblematischer‹ medialer Gebrauch ist eben noch längst nicht gegeben. Es scheint daher einiges dafür zu sprechen – und auch dieses Argument wurde in den vergangenen zwei Jahren häufig gemacht (vgl. CHESHER / ALBARRÁN-TORRES 2023: 3; Übers. L.W.) – dass KI-Bilder sich derzeit noch in einer Phase der »säkularen Magie« befinden; des stets mitkommunizierten *Spektakels* und der Verblüffung; vielleicht nicht unähnlich der frühen Fotografie als Bühnenmagie, bevor der Zauber verflogen war und sich stabile Verwendungsweisen etabliert hatten. Erst daraufhin, so das Argument, konnte Fotografie zum Massenmedium werden, an das schließlich eigene mediale Protokolle (wie die Portrait-Fotografie bis hin zum Selfie) angeschlossen werden konnten.

Der mediale ›Ort‹ der KI-Bildzirkulation – zumindest, wo es eben *nicht* um Täuschung, um ein bewusstes oder unbewusstes Missverstehen von KI-Bildern als Fotos oder materiell gefertigten Artefakten, geht, sondern um eine Verwendung *als KI-Bilder* – liegt derzeit vor allem in der Social Media-Kommunikation und in Fankulturen (vgl. LAMERICHS 2023), wo Prompts wie Kochrezepte oder Zauberformeln ausgetauscht oder sorgsam geheim gehalten werden. Die paradigmatische Verwendungsweise von KI-Bildern ist hier die Meme-Kommunikation. Die KI-Genese wird hier einerseits durch multimodale Paratexte hergestellt (Kommunikation über Prompt, Modell und Plattform), andererseits durch die erkennbare Zugehörigkeit eines Bildes zu einer Serie und schließlich durch den Publikationsort in ausgewiesenen Reddit- oder Facebook-Gruppen wie »AI Art Universe«.

Auf dem ›hypermediacy‹-basierten Pol des Spektakels geht es dann um Mashups und Remixes existierender Contents und Styles (vgl. auch BOLTER 2023). Typisch scheinen mir hier etwa die Re-Imaginationen bekannter Filme durch andere Regisseure mit wiedererkennbarer Ästhetik, allen voran etwa Wes Anderson, der zu seinem Unmut zum favorisierten ›Style‹-Spender vieler KI-Bildschöpfungen geworden ist. Dass diesen Remix- und Mashup-Praxen auch ein subversives und kreatives Potenzial zukommt, liegt auf der Hand. »Harry Potter as a Bollywood Movie« (ET ONLINE 2023: n.pag.) lässt sich doch sicher als neue, leicht zugängliche Form von visueller Fan Fiction deuten, die einen eurozentrischen Blick auf bestehende Figuren umdeuten möchte. Dass dabei vielleicht zugleich äußerst stereotype Bilder von typischen »Bollywood«-Filmen fortgeschrieben und festgeschrieben werden (wer schließlich hat in welchen Datenbanken hinterlegt, welche Bilder als »Bollywood« verschlagwortet sind?), ist sicherlich auch hier die andere Seite der Medaille.

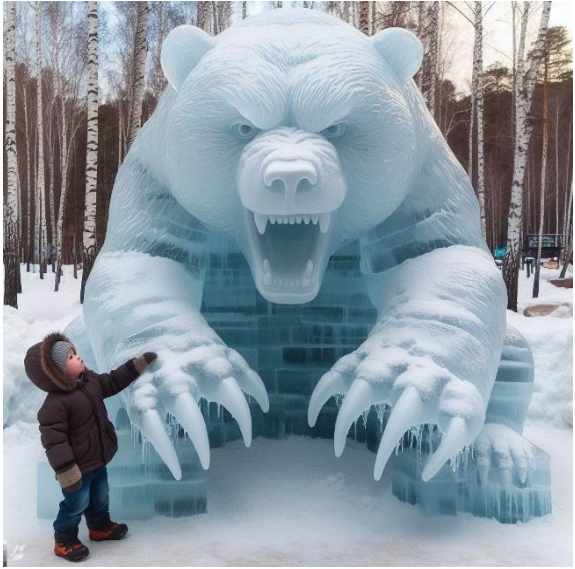

Abb. 5: »My child did this. Please don't judge harsh of his art, he's only 3« (ALTON 2024)

Auf dem ›immediacy‹-basierten Pol des Spektakels scheinen mir solche Meme-Serien typisch, die zwar Fotos (oder Foto-Ästhetik) remediieren, die inszenierte ›Realität‹ aber zugleich über ihren Inhalt ironisieren. Ganze Serien von Memes etwa präsentieren scheinbar menschliche Protagonist*innen, die stolz irgendwelche materiellen, scheinbar handgefertigten Kreationen vorführen: »The little Max has drawn a picture. Let's not criticise strictly and let's encourage him!« (sowohl ›Max‹ als auch sein Bild sind KI-generiert, vgl. GRENHOLM 2024), »My child did this. Please don't judge harsh of his art, he's only 3« (das KI-Bild zeigt eine völlig überbordende, meterhohe Eisskulptur, vgl. ALTON 2024, Abb. 5) oder »This young man built a BMW car using only Bricks. Make him trend« (ein ›afrikanischer Junge‹ neben einem perfektem BMW, lediglich aus ›Stein‹ gefertigt, vgl. BROWN 2024).[7] Spannend an diesen Meme-Serien ist nicht nur, dass sie stets mit der doppelten Adressierung in den Kommentarspalten spielen, in denen stets jemand daran gutgläubig zweifelt, dass diese ›menschlichen Leistungen‹ tatsächlich echt seien – obgleich bereits Gruppennamen wie »AI Art Universe« keinerlei Zweifel aufgeben sollten; andererseits scheint mir das inszenierte Spektakel um die angeblichen menschlichen Leistungen eine merkwürdige, inszenierte Verschiebung des Staunens darüber zu bieten, eben diese neuen, durchaus beeindruckenden KI-Bilder (mit welchen Prompts? Mit welchem Modell?) hergestellt zu haben. Beide Fälle, das hypermediacy-Spektakel und das immediacy-Spektakel, spielen in jedem Fall durchaus mit der Verblüffung um die bildlichen Resultate.

### 1.6 Schlussüberlegungen – Verlaufslinien aus der primären Intermedialität heraus

Da der Einsatz von KI-Bildern potenziell zu vielfältig ist und ohne Frage noch weiter zunehmen wird, bleibt die Frage nach möglichen Verlaufslinien aus der primären Intermedialität hinaus – auch jenseits des engen Anwendungsfeldes von Social Media-Memes. Erneut wären hier sicherlich Vergleiche mit der frühen Fotografie oder dem Bewegtfilm aufschlussreich, die sich Schritt für Schritt von ›säkularer Magie‹ zu einer

---

[7] All diese Memes gehen wohl auf Foto-Posts eines tatsächlichen Handwerker-Künstlers namens Michael Jones zurück, dessen Motorsägen-Skulpturen seit Dezember 2023 in großer Zahl von KI-Bildgeneratoren plagiiert werden, vgl. KOEBLER 2023.

Medienform mit angeschlossenen Institutionen, Produktionsrollen, Distributions- und Rezeptionsräumen und schließlich eigenen inhaltlichen Formen und Genres entwickelt haben. Damit wäre ein erster Weg vorgezeichnet, der für KI-Bilder ebenfalls bereits von Praktiker*innen und Produzent*innen angedacht worden ist: Die Russo-Brothers, bekannt etwa als Regisseure der *Avengers: Endgame* and *Avengers: Infinity War*-Filme, spekulierten bereits im April 2023 öffentlich darüber, dass innerhalb von zwei Jahren ein neues KI-Medium entstehen könnte, dass konzeptuell *zwischen* Filmen und Videospielen läge: »You could walk into your house and save the AI on your streaming platform. ›Hey, I want a movie starring my photoreal avatar and Marilyn Monroe's photoreal avatar. I want it to be a rom-com because I've had a rough day,‹ and it renders a very competent story with dialogue that mimics your voice« (zit. nach SHARF 2023: n.pag.). Solche Visionen sind, wie Leschke ebenfalls deutlich gemacht hat, ebenfalls noch fest im primären Intermedialitätsdiskurs zuhause, verbinden sie doch technische Prognosen mit »sozialutopischen Versprechen« (LESCHKE 2004: 61). Damit aber wäre eine Einheit von Technik, Verwendungsweise und Inhalten, ein *qualifiziertes* Medium bzw. ein *konventionell distinktes* Medium hergestellt.

    Ein anderer Weg bestünde gerade in der gegenläufigen Richtung: Dem Auflösen der intermedialen Differenz, dem nahtlosen *Aufgehen* von KI-Bildern in der Medienkonvergenz. Wenn *jedes* Bild in Zukunft prinzipiell *immer* zumindest teilweise von generativen KI-Netzwerken erzeugt oder modifiziert sein kann – und dafür spricht vor allem Adobes Integration von Firefly in jede Funktion von Photoshop – dann könnte diese Auszeichnung zukünftig völlig unentscheidbar, nachgeordnet und letztlich irrelevant werden; ebenso wie es schon jetzt einigermaßen bedeutungslos geworden ist, von ›digitalen‹ Bildern zu sprechen. Sowohl Handy- als auch Analogfotografie ist doch wohl irgendwann durch Photoshop gelaufen oder wurde digital versandt. So, wie wir also bereits in der ›post-digitalen‹ Gesellschaft angekommen sind (vgl. JUNG / SACHS-HOMBACH / WILDE 2021), so könnten wir nun bereits mit einem Fuß in einer »postartifiziellen« Gesellschaft stehen, wie man mit Bajohr (2023a: 37) sagen könnte. Jedes denkbare Bild, in jedem denkbaren Stil, könnte zukünftig immer *mehr oder weniger* KI-modifiziert sein, ohne dass dies eine kategoriale Unterscheidung bleibt – von der die gesamte Frage nach Intermedialitätsverhältnissen doch ausgeht.

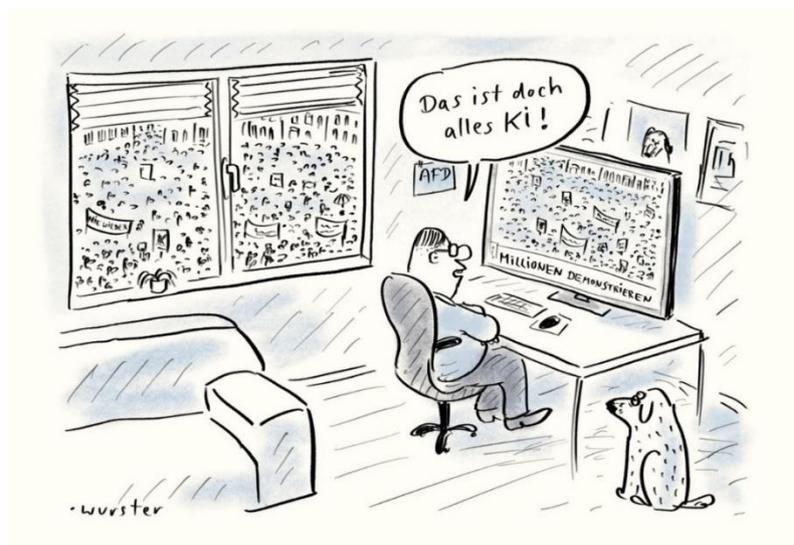

Abb. 6: Miriam Wursters »Proteste gegen Rechtsextremismus« (WURSTER 2024)

Wenn beide Fluchtpunkte der Intermedialitätsdiskurse im Moment Spekulation bleiben müssen, so lässt sich als Schlussbeobachtung lediglich ein Hinweis auf Ebene einer möglichen generellen Medienontologie festhalten (vgl. Leschke 2003: 161-236). Ebenso wie der Computer bzw. die Digitalisierung nicht lediglich ein weiteres, additiv hinzugefügtes neues Medium darstellte, sondern die Medienökologie generell transformiert hat – ohne dabei freilich die als relevant erachteten sekundären intermedialen Interferenzen zum Verschwinden zu bringen, sie sicherlich jedoch zu verschieben – ebenso scheinen KI-Bilder heute *generell* unser Verständnis von Bildlichkeiten zu verändern. Wie ein Cartoon von Miriam Wurster aus der *Süddeutschen Zeitung* vom Januar 2024 persifliert (vgl. WURSTER 2024, Abb. 6), schreibt sich derzeit eine generelle Hermeneutik des Verdachts in jedes einzelne vorliegende und zukünftige Bild ein. Wir können oder müssen nun plötzlich überall und jederzeit KI-Produktion vermuten und unterstellen. Von den tatsächlich vorhandenen technologischen Möglichkeiten scheint dies abermals nur milde bedingt, denn, natürlich, seit mehreren Jahrzehnten ließe sich derselbe Verdacht bereits hinsichtlich möglichen Photoshop-Manipulationen aussprechen. Dass dies in der Regel aber nicht grundsätzlich der Fall war, dass der visuelle Evidenzeffekt entgegen aller Empirie weitgehend ungebrochen blieb und eben nur in begründeten Einzelfällen problematisch wurde, dies mag sich nun durch die niederschwellige und billige Zugänglichkeit und die schier grenzenlose Flut von KI-Bildern ändern. Wie unsere sozial-kulturellen Protokolle der Bildproduktion, -distribution und -rezeption mittel- und längerfristig darauf reagieren werden, bleibt entsprechend spannend. Die Medien- und Bildtheorie wird sicherlich noch viele Gelegenheiten haben, sich daran abzuarbeiten.

Trondheim, 10. April 2024

Preprint: Angenommen zur Publikation in Marcel Lemmes, Stephan Packard und Klaus Sachs-Hombach (Hrsg.): *Zukunft der Bilder: Herausforderungen der Bildwissenschaft*. Köln [Herbert von Halem] 2025/in Vorbereitung.

**Literatur:**

Preprint: Angenommen zur Publikation in Marcel Lemmes, Stephan Packard und Klaus Sachs-Hombach (Hrsg.): *Zukunft der Bilder: Herausforderungen der Bildwissenschaft*. Köln [Herbert von Halem] 2025/in Vorbereitung.

Preprint: Angenommen zur Publikation in Marcel Lemmes, Stephan Packard und Klaus Sachs-Hombach (Hrsg.): *Zukunft der Bilder: Herausforderungen der Bildwissenschaft*. Köln [Herbert von Halem] 2025/in Vorbereitung.THON, JAN-NOËL: Mediality. In: Marie-Laure Ryan; Benjamin J. Robertson; Lori Emerson (Hg.): *The Johns Hopkins Guide to Digital Media*. Baltimore [Johns Hopkins University Press] 2014, S. 334-336

WASIELEWSKI, AMANDA: »Midjourney Can't Count«: Questions of Representation and Meaning for Text-to-Image Generators. In: *Generative Imagery: Towards a ›New Paradigm‹ of Machine Learning-Based Image Production, special-themed issue of IMAGE: The Interdisciplinary Journal of Image Sciences, 37, 2023,* S. 71-82

WILDE, LUKAS R.A.: Generative Imagery as Media Form and Research Field: Introduction to a New Paradigm. In: *Generative Imagery: Towards a ›New Paradigm‹ of Machine Learning-Based Image Production*, special-themed issue of *IMAGE: The Interdisciplinary Journal of Image Sciences,* 37(1), 2023, S. 6-33

WURSTER, MIRIAM: Proteste gegen Rechtsextremismus. In: *Süddeutsche Zeitung*. 23. Januar 2024. https://www.sueddeutsche.de/meinung/wurster-proteste-gegen-rechtsextremismus-1.6337465 [*letzter Zugriff: 12.04.2024*]
**Abbildungen:**

Abb. 1: Boris Eldagsens »Pseudomnesia: The Electrician« (vgl. GLYNN 2023)

Abb. 2: Der berüchtigte »Baby Peacock«, der in Wirklichkeit natürlich nicht so aussieht (vgl. LAMAGDELEINE 2023)

Abb. 3: »fluffy glamour glow« (MEYER 2023a)

Abb. 4: »This was posted unironically, what a world 😂« (SOUTHEN 2024)

Abb. 5: »My child did this. Please don't judge harsh of his art, he's only 3« (ALTON 2024)

Abb. 6: Miriam Wursters »Proteste gegen Rechtsextremismus« (WURSTER 2024)

**25** | 25